\documentclass[
aps,%
12pt,%
final,%
notitlepage,%
oneside,%
onecolumn,%
nobibnotes,%
nofootinbib,
superscriptaddress,%
noshowpacs,%
centertags]%
{revtex4}
\usepackage{graphicx}
\usepackage[cp866]{inputenc}
\usepackage[T2A]{fontenc}
\usepackage[english,russian]{babel}

\begin{document}
\selectlanguage{english} 
\title{Theoretical analysis of the
astrophysical S- factor for the $\alpha+d\rightarrow^{6}Li+\gamma $
capture reaction in the two body model}
\author{E.~M.~ Tursunov}
%
\author{S.~ A.~ Turakulov}
\affiliation{Institute of Nuclear Physics, Uzbekistan Academy of
Sciences, 100214, Ulugbek, Tashkent, Uzbekistan}
\author{P.~ Descouvemont}
\affiliation{Physique Nucl\'eaire Th\'eorique et Physique
Math\'ematique, C.P. 229, Universit\'e Libre de Bruxelles, B1050
Brussels, Belgium}

\begin{abstract}
Theoretical estimations for the astrophysical S-factor and the
$d(\alpha,\gamma)^{6}Li$ reaction rates are obtained on the base of
the two-body model with the $\alpha-d$ potential of a simple
Gaussian form, which describes correctly the phase-shifts in the S-,
P-, and D-waves, the binding energy and the asymptotic normalization
constant in the final S-state. Wave functions of the bound and
continuum states are calculated by using the Numerov algorithm of a
high accuracy. A good convergence of the results  for the E1- and
E2- components of the transition is shown when increasing the upper
limit of effective integrals up to 40 fm. The obtained results for
the S-factor and reaction rates in the temperature interval $ 10^{6}
K \leq T \leq 10^{10} K $ are in a good agreement with the results
of Ref. A.M.~ Mukhamedzhanov, et.al., Phys. Rev. {\bf C 83}, 055805
(2011), where the authors used the known asymptotical form of wave
function at low energies and a complicated potential at higher
energies.
\end{abstract}
\maketitle

\newcommand {\nc} {\newcommand}
\nc {\beq} {\begin{eqnarray}} \nc {\eol} {\nonumber \\} \nc {\eeq}
{\end{eqnarray}} \nc {\eeqn} [1] {\label{#1} \end{eqnarray}} \nc
{\eoln} [1] {\label{#1} \\} \nc {\ve} [1] {\mbox{\boldmath $#1$}}
\nc {\vS} {\mbox{$\ve{S}$}} \nc {\cA} {\mbox{$\cal{A}$}} \nc
{\dem} {\mbox{$\frac{1}{2}$}} \nc {\arrow} [2]
{\mbox{$\mathop{\rightarrow}\limits_{#1 \rightarrow #2}$}}



\section{Introduction}

\par It is well known that the $^6$Li nuclei have been formed mainly as a result of
the Bing Bang through the capture reaction
\begin{eqnarray} \label{1}
\alpha+d\rightarrow ^6Li+\gamma
\end{eqnarray}
at low energies  $50 \le E_{cm} \le 400$ keV \cite{ser04}. This
process was studied in details by the experimental groups at
energies around the $3^+$  resonance with  $E_{cm}=$0.711 MeV and at
higher energies \cite{mohr94,robe81}. However, at low energies the
obtaining an information on the cross section of the process from
the analysis of the experimental data met insuperable difficulties
\cite{kien91,hamm10}. In the recent work \cite{hamm10} the break up
process of the $^6Li$ nucleus in the field of heavy ion $^{208} Pb$
was studied with the aim to extract data on the cross section of the
backward process at astrophysical energies in laboratory conditions.
Unfortunately, a dominance of the nuclear break up over the Coulomb
process did not allow to realize this idea.

\par From the theoretical point of view, the synthesis reaction of the $^6Li$ nucleus
was studied in microscopic and macroscopic potential models
\cite{lang86,mohr94,type97,desc98,dub95}, and also in the "ab
initio" calculations \cite{noll01}. In recent work \cite{mukh11} it
was strongly argued that the two-body model of the synthesis process
$^2H(\alpha,\gamma)^6Li$ should be based on $\alpha-d$ potentials,
which describes the phase shifts in partial waves and additionally
reproduces the binding energy $E_b=1.474$ MeV and the asymptotic
normalization coefficient (ANC) in the S-wave of the $\alpha+d$
bound state. In Ref. \cite{blok93} it was shown that ANC can be
extracted from the analysis of the experimental data on the
$\alpha-d$ elastic scattering and it's value was established with
some error bars as $C_{\alpha d}=2.30 \pm 0.12$ fm$^{-1/2}$. At the
end, in above mentioned work \cite{mukh11}, the theoretical
estimations of the astrophysical S-factor and corresponding reaction
rates of the synthesis reaction $^2H(\alpha,\gamma)^6Li$ at low
energies $E \le 300$ keV have been obtained with the help of the
asymptotical form of the bound state wave function of the $\alpha+d$
system   $C_{\alpha d}W_{-\eta,1/2}(2 k_{\alpha
 d}r)/r$. And at higher energies, where the internal structure of the wave function
 is important, the calculations have been done with a potential of
 a complicated form, which is phase equivalent to the original Wood-Saxon
 potential from Ref. \cite{hamm10}, and reproduces the binding energy and ANC
 of the  $\alpha+d$ system with $C_{\alpha d}=2.28$ fm$^{-1/2}$. At the same
 time the initial potential overestimates the ANC by 0.42 fm$^{-1/2}$. The phase equivalent
 potential was built with the help of a complicated integro-differential transformations.
 Since the astrophysical S-factor is proportional to the square of ANC, then
 it's value decreases by about 40 percent comparing to the initial value, obtained by the Wood-Saxon potential.
 One can ask here a question: is it possible to reproduce the results of
 Ref. \cite{mukh11} on the base of a simple local $\alpha+d$ potential, which correctly describes
 phase shifts in partial waves and bound state properties, i.e. binding energy and ANC?

\par On the other hand, in Ref. \cite{dub10} a central potential of
the  Gaussian form with additional Coulomb interaction, containing
Pauli forbidden states in the partial S- and P-waves have been used
for the estimation of the astrophysical S-factor of the capture
process and the estimation  1.67 eV mbn  has been obtained in the
energy region around 5-10 keV. We note that the estimation of ANC
$C_{\alpha d}=2.53$ fm$^{-1/2}$ for the Gaussian potential
overestimates the corresponding value from Ref. \cite{mukh11} by
0.25 fm$^{-1/2}$. At the same time, the Gaussian potentials
reproduce the phase shift of the  $\alpha-d$ elastic scattering in
the S, P, D -waves up the energy value $E=9$ MeV and the binding
energy of the $^6$Li nucleus. It is important to note that for the
calculation of the bound state wave function of the $ \alpha+d$
system  an expansion over 10 Gaussians have been used, which does
not describe well the asymptotics  even at distances about 10-15 fm.
Therefore the authors of this work, as well as the authors of Ref.
\cite{mukh11} have used the known asymptotic form of the wave
function at large distances for the calculations of the
characteristics of the above capture process.

\par The aim of current work is a detailed theoretical analysis
of the astrophysical S-factor and corresponding reaction rates in
the two-body model on the base of the $\alpha-d$ potential of a
simple Gaussian form, which describes correctly the phase shifts in
the partial $^3S_1, \, ^3P_0, \, ^3P_1, \, ^3P_2, \, ^3D_1, \,
^3D_2, \, ^3D_3 $ waves, and the binding energy and ANC of the bound
state in the S-wave. In our work we are based on the $\alpha-d$
potential from Ref. \cite{dub94}, but for the calculation of the
wave functions we use the Numerov algorithm, which has an accuracy
of order $O(h^6)$ \cite{numerov}. This high accuracy allows one to
obtain wave functions, which are well consistent with the known
asymptotics in the each partial wave. Further we will show that the
S-wave potential can be modified in such a way, that reproduce the
ANC, while the binding energy remains unchanged. At the same time,
the description of the phase shifts is improved and the theoretical
phase shift is more consistent with the late data \cite{jen83} than
with the old data \cite{mc67,grue75}.
\par  In Section 2 we give the used model, in Section 3 we give numerical results and
 conclusions are given in the last Section.

\section{Model}

\subsection{Wave functions}

 The wave function of the initial $\alpha+d$ scattering states in the  $^3P_0,  \,
^3P_1, \, ^3P_2, \, ^3D_1, \, ^3D_2, \, ^3D_3 $  partial waves and
the final $^3S_1$ bound state are found as solutions of the two-body
radial Schr\"{o}dinger equation
\begin{eqnarray}
\left[-\frac{\hbar^2}{2\mu}\left(\frac{d^2}{dr^2}-\frac{l(l+1)}{r^2}\right)+V^{J
ls}(r)\right] \phi(r)=E \phi(r),
\end{eqnarray}
where $V^{J ls}(r)$ is a $\alpha+d$ two-body potential in the
partial wave with the orbital momentum $l$, spin $s$ and total
momentum $J$. For the solution of the equation we use  the Numerov
algorithm. As we see below, the calculated wave functions are of a
high accuracy, that is necessary when applying to the estimations of
the characteristics of the astrophysical capture reaction
$^2H(\alpha,\gamma)^6Li$.

 A radial scattering wave function $u_E(r)$ is normalized  with the help of the
 asymptotical relation
  \beq u_E(r) \arrow{R}{\infty} \cos\delta_l(E) F_l (kr)
+ \sin\delta_l(E) G_l(kr), \label{eq220} \eeq where  $k$ is the wave
number of the relative motion, $F_l$ and  $G_l$ are Coulomb
functions and  $\delta_l(E)$ is the phase shift in the partial wave.

\subsection{Cross section of the capture process and the astrophysical S-factor}

\par The differential cross section of the synthesis process
$^2H(\alpha,\gamma)^6Li$ in the two body model in the temperature
interval $ 10^{6} K \leq T \leq 10^{10} K $ is expressed as
\cite{Angulo}
\begin{displaymath}
\sigma(E)=\sum_{J_f \lambda}\sigma_{J_f \lambda}(E),
\end{displaymath}
\begin{eqnarray}
\sigma_{J_f \lambda}(E)= \frac{8 \pi e^2}{\hbar v q^2} \left[Z_1
\left( \frac{A_2}{A} \right)^{\lambda}+Z_2 \left(\frac{-A_1}{A}
\right)^{\lambda} \right]^2 C^2(S_{J_f}) \\ \nonumber \times
\sum_{J_i,S,l_i}\frac{(k_{\gamma})^{2\lambda+1}}{[(2\lambda+1)\textrm{!!}]^2}\frac{(\lambda+1)(2\lambda+1)}{\lambda}\\
\nonumber \times \frac{(2 l_i +1)(2
l_f+1)(2J_f+1)}{(2S_1+1)(2S_2+1)} \left( \begin{array}{ccc}
l_{f} & \lambda & l_{i} \\
0 & 0 & 0
\end{array} \right)^{2}\\ \nonumber
\times (2J_i+1) \left\{ \begin{array}{ccc}
J_{i} & l_{i} & S \\
l_{f} & J_{f} & \lambda
\end{array} \right\} ^{2} \left(\int^{\infty}_{0} u_{i}(r)r^{\lambda} u_{f} (r) dr
\right)^{2},
\end{eqnarray}
where  $u_{i}$ and $u_{f}$ are the wave functions of the initial
scattering and final bound states, $k_{\gamma}$ is the photon
quantum number, $l_{i},J_{i},l_{f},J_{f}$ are the orbital and total
momenta of the initial and final states, respectively, $\lambda$ is
a multiplicity of the electric (E) transition,
 $S_1$, $S_2$ are spins of the clusters, $A=A_1+A_2$, $A_1$, $A_2$, $Z_1$,
 $Z_2$ are experimental mass and charge values of the cluster in the entrance
 channel. As was argued in Ref.\cite{mukh11}, a value of the spectroscopic
 factor $C^2 (S_{J_f})=1$, when using the two-body potentials, which reproduce
 correctly the phase shifts in partial waves.

The astrophysical $S$-factor of the process is expressed through the
cross section as \cite{Fowler}
\begin{eqnarray}
S(E)=E \, \, \sigma(E) \exp(2 \pi \eta),
\end{eqnarray}
where  $\eta$ is the Coulomb parameter.

The reaction rate $N_{a}(\sigma v)$ is estimated with the help of
well known expression \cite{Fowler, Angulo}
\begin{eqnarray}
N_{a}(\sigma v)=N_{A} \frac{(8/\pi)^{1/2}}{\mu^{1/2}(k_{B}T)^{3/2}}
\int^{\infty}_{0} \sigma(E) E \exp(-E/k_{B}T) d E
\end{eqnarray}
where $k_{B}$ is the Boltsman constant, $T$ is the temperature,
$N_{A}=6.0221\times10^{23}\text{mol}^{-1}$  the Avogadro number.
When $k_{B}T$ is expressed in MeVs, it is convenient to introduce a
variable $T_9$ for the temperature in the units of $10^9\, K $ with
the help $k_{B}T=T_{9}/11.605$ MeV which varies in the interval
$0.001\leq T_{9} \leq 10 $. After the substitution of the specified
values we have next expression for the integral:
\begin{eqnarray}
\label{rate}
 N_{a}(\sigma v)=3.7313 \times 10^{10}A^{-1/2}T_{9}^{-3/2}
\int^{ \infty}_{0} \sigma(E) E \exp(-11.605E/T_{9}) d E.
\end{eqnarray}

\section{Numerical results}

\par For the solution of the Schr\"{o}dinger equation in the entrance and exit channels
 we use the two-body $\alpha-d$ central potentials of the Gaussian
 form wit the corresponding Coulomb interaction from Ref. \cite{dub10}
 with  $\hbar^2/2 m_N=20.7343$ MeV fm$^2$. The experimental mass values are chosen
 as in the indicated work \cite{dub10}: $A_1=$ 2.013553212724 a.u.m. and
$A_2=$4.001506179127 a.u.m.
 The potentials in the partial waves $^3S_1, \, ^3P_0, \, ^3P_1,
\, ^3P_2 $ contain additional Pauli forbidden states which have a
microscopical background, but there are no such states in the
$^3D_1, \, ^3D_2, \, ^3D_3 $ channels. As was noted above, numerical
solution of the Schr\"{o}dinger equation was obtained with the use
of the Numerov algorithm on the base of the Newton-Rapson method.
The step is fixed as $h=$0.05 fm, and the number of the mesh points
are varied from $N=$200 up to $N=$2000 for the check of convergence
of numerical results.

\par The initial Gaussian potential in the $^3S_1$ wave
 $V_D(r) = -76.12 \exp(- 0.2 r^2)$ MeV \cite{dub94}
describes well the phase shifts of the $\alpha-d$ scattering (see
Fig. 1), however a consistence with the later data from the work
\cite{jen83} is not so well.

 On Fig. 2 we give a description of the asymptotics of the bound state wave function
 of the $\alpha+d$ system. As can be seen from the figure, the calculated wave function
 on the base of the Numerov algorithm is well consistent with the asymptotics
even with the number of mesh points $N=$200, which corresponds to
$R_{max}=$ 10 fm. However, the initial potential $V_D(r)$ yields the
estimation for the ANC of the $\alpha+d$ system with
 $C_{\alpha d}=2.53$ fm$^{-1/2}$, which is larger than the estimation from Ref.
\cite{blok93} extracted from the experimental data on the $\alpha+d$
scattering by about 0.23 fm$^{-1/2}$. Therefore, in accordance with
the ideology of Ref.\cite{mukh11}, we slightly modify the initial
potential in such a way, that the resulting potential reproduces the
correct value of ANC. On Fig.1 we also show the description of phase
shifts in the S-wave with the modified potential $V_M(r)=-92.44
\exp(- 0.25 r^2) $ MeV, which are well consistent with the later
data from Ref. \cite{jen83}. The same potential reproduces the
empirical ANC  with  $C_{\alpha d}=2.31$ fm$^{-1/2}$ (see Fig. 2).

\par For the examination of the convergence of the theoretical results
for the astrophysical S-factor and the differential cross section of
the capture process $^2H(\alpha,\gamma)^6Li$, it is convenient to
introduce an effective integral
\begin{eqnarray}
I_{eff}(R,\lambda)=  \left[Z_1 \left( \frac{A_2}{A}
\right)^{\lambda}+Z_2
\left(\frac{-A_1}{A} \right)^{\lambda} \right] \\
\nonumber \times
\frac{1}{[(2\lambda+1)\textrm{!!}]}\sqrt{\frac{(\lambda+1)(2\lambda+1)}{\lambda}}
 \int^{R}_{0} u_{i}(r)(k_{\gamma}r)^{\lambda} u_{f} (r) dr,
\end{eqnarray}
the square of which is contained in the expression for the cross
section at  $R=\infty$. Thus, we can check a behavior of the
effective integral with increasing  $R$ at several energy values and
with the fixed entrance channel and multiplicity  $\lambda$ of the
electric transition.

 On Fig.3 we show effective integrals for the entrance channels  $^3P_2 $
 and $^3D_3 $, which give maximal contributions to the E1 and E2
 transitions, correspondingly, at energies E=0.1, 0.5 and 1 MeV. From the figure
 one can see that the effective integrals for the E1 transition at all the energy
 values converge faster than the effective integrals for the E2 transition.
 For the E2 transition the convergence is achieved at 35-40 fm. Here it is important
 to note that at higher energies (for example at 1 MeV) the integrals for the E2
 transition change the sign at 10-15 fm, which occurs due to the mutual cancellation
  of the internal and asymptotic parts of the transition matrix elements.
At the same time, this situation is due to the presence of the extra
Pauli forbidden state in the S-wave of the  $\alpha+d$ two body
system, that gives a node in the internal part of the ground state
wave function. The mutual cancellation of the matrix elements
allowed to reproduce data on the beta-transition of the $^6He$ halo
nucleus into the $\alpha+d$ two-body continuum channel \cite{tur06}.
At the end, in such cases the main contribution to the effective
integrals comes from the asymptotic parts of the wave functions. For
the E1 transition the wave functions of the entrance (P-wave) and
exit (S-wave) channels have nodes due-to the Pauli forbidden states
approximately at the same position, hence their product keeps the
sign up to large distances. Therefore here one can not see any
cancellation effects.

On Fig.4 the contributions from the partial waves to the
E1-component of the astrophysical S-factor for the
$\alpha+d\rightarrow^{6}Li+\gamma $ synthesis reaction calculated
with the potential $V_M$ are demonstrated. As can be seen from the
figure, the partial wave $^3P_2 $ in the entrance channel yields the
dominant contribution to the E1-transition.

On Fig.5 we show the corresponding contributions from the partial
waves to the E2-component of the astrophysical S-factor, calculated
with the potential $V_M$. In this case, the dominant contribution to
the process comes from the $^3D_3 $ entrance channel in the energy
interval up to the resonance region, and the maximal contribution
behind the resonance comes from the $^3D_2 $ entrance channel.

\par For the visual demonstration of the convergence of the obtained theoretical results,
on Fig.6 we show the contributions of the E=E1+E2 transitions to the
astrophysical S-factor for the capture reaction
$^2H(\alpha,\gamma)^6Li$, estimated with the potential $V_M$ at
several sets of mesh points with N=200, 400, 600, 800-2000 in the
wide energy region. Since the step is fixed  as $h=$0.05 fm, the
respective upper limits of the integrals are 10, 20, 30, 40-100 fm.
From the figure one can see that for the complete convergence it is
necessary to choose the integral upper limit not less than 40 fm.

\par The contributions of the E1,E2, E1+E2 transitions to the astrophysical
S-factor for the capture reaction $\alpha+d\rightarrow^{6}Li+\gamma
$, estimated with the potentials $V_M$ and $V_D$ in comparison with
the experimental data from Refs.\cite{mohr94,robe81,kien91,iga00}
are shown on Fig.7. From the figure one can see that at low energies
the main contribution to the astrophysical S-factor comes from the
E1-component, and at energies around and behind the $^3D_3$
resonance region the contribution of the E2 component is dominant.
We note also, that the initial potential $V_D$ from Ref.\cite{dub94}
overestimates the experimental data in the region behind the
resonance. But at low energies up to the resonance region the
experimental data, as was noted above, are not well defined.
Therefore it is too early to make a conclusion about the level of
description of the data by the theoretical models at the low
astrophysical energies. However, the theoretical results, obtained
with the modified potential $V_M$ are very consistent with the
results of Ref. \cite{mukh11}.

\par In the Table 1 we give our theoretical estimations for the reaction rates
of the process  $d(\alpha,\gamma)^{6}Li$ in the temperature interval
$ 10^{6} K \leq T \leq 10^{10} K $ ($ 0.001\leq T_{9} \leq 10 $) in
comparison with the results of Refs. \cite{hamm10,mukh11}. In the
second column we give "the most effective energy"  $E_0$, which
gives the maximum to the integrant in Eq.(\ref{rate}). It is
expressed as \cite{Angulo}
\begin{eqnarray}
E_0=\left( \frac{\mu}{2} \right)^{1/3} \left( \frac{\pi e^2 Z_1Z_2
k_B T}{\hbar} \right)^{2/3}= 0.122 \, (Z_1^2 Z_2^2 A)^{1/3} T_9
^{2/3} \, \textrm {MeV,}
\end{eqnarray}

where $\mu$ is the reduced mass of the two particles.
 From the table one can find a good agreement of our results,
obtained by using the modified potential $V_M$, with the results of
Ref.~\cite{mukh11}. However, our estimations are slightly lower than
the results of the mentioned work. Probably, this is connected with
the different potential choices and also with the fact that in
Ref.~\cite{mukh11} the asymptotical form of the $\alpha+d$ two-body
wave function has been used for the estimation of the reaction rates
at energies up to 350 keV.

\section{Conclusions}

\par The astrophysical S-factor and corresponding reaction rates
for the process $d(\alpha,\gamma)^{6}Li$ have been estimated on the
base of the two body model with the $\alpha-d$ potentials of a
simple Gaussian form, which describe correctly the phase shifts in
the partial waves $^3S_1, \, ^3P_0,  \, ^3P_1, \, ^3P_2, \, ^3D_1,
\, ^3D_2, \, ^3D_3 $, and also the binding energy and ANC of the
bound state in the S-wave. By modifying the S-wave potential from
Ref.~\cite{dub94}, we obtained a better description of the phase
shifts and ANC, while keeping the binding energy of the $^{6}Li$
nucleus unchanged. For the calculations of the wave functions in the
bound and continuum channels we have used the Numerov algorithm,
which is of a high accuracy and  yields the correct asymptotics of
the wave function in the each partial wave. It was shown that a good
convergence of the estimations for the contributions of the E1- and
E2- transitions to the astrophysical S-factor is obtained when the
upper limit of the integrals are extended up to 40 fm.

 The theoretical results for the astrophysical S-factor and reaction rates
 of the process  $d(\alpha,\gamma)^{6}Li$  in the temperature interval
  $10^{6} K \leq T \leq 10^{10} K $ ($ 0.001\leq T_{9} \leq 10 $)
 are in a good agreement with the results of Ref.~\cite{mukh11}, obtained by using
 known asymptotical form of the wave function at low energies and a complicated
two-body potential at higher energies.

 E.M.T. thanks Prof. L.D. Blokhintsev and Prof. B.F. Irgaziev for useful comments.

\newpage
%
\begin{figure}
\includegraphics[width=20cm]{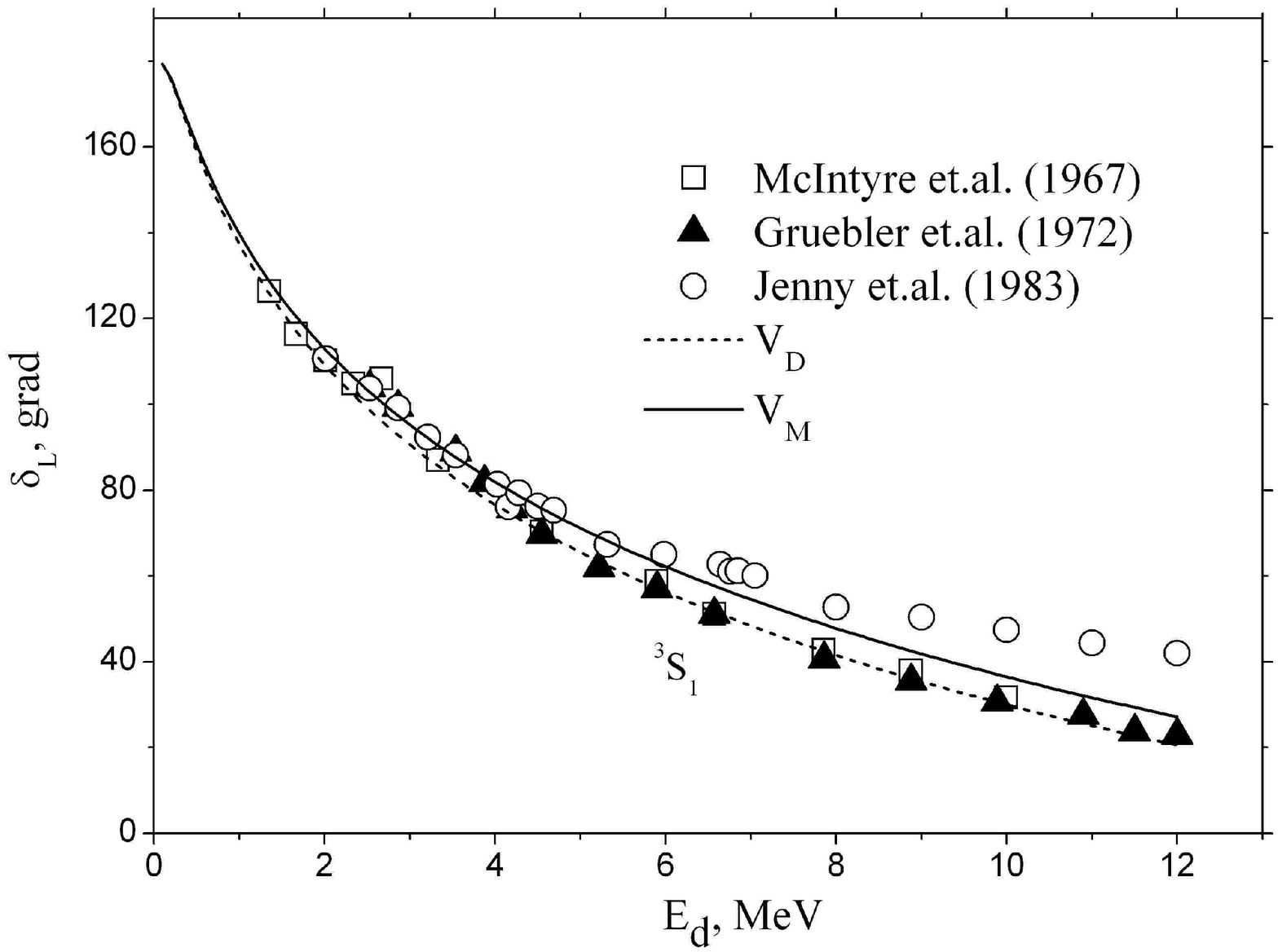}
\caption{Phase shifts of the $\alpha+d$ elastic scattering in the
S-wave with potentials $V_D$ and  $V_M$ in comparison with the
experimental data from Refs.~\cite{mc67,grue75,jen83}.}
\end{figure}
\newpage
%
\begin{figure}
\includegraphics[width=20cm]{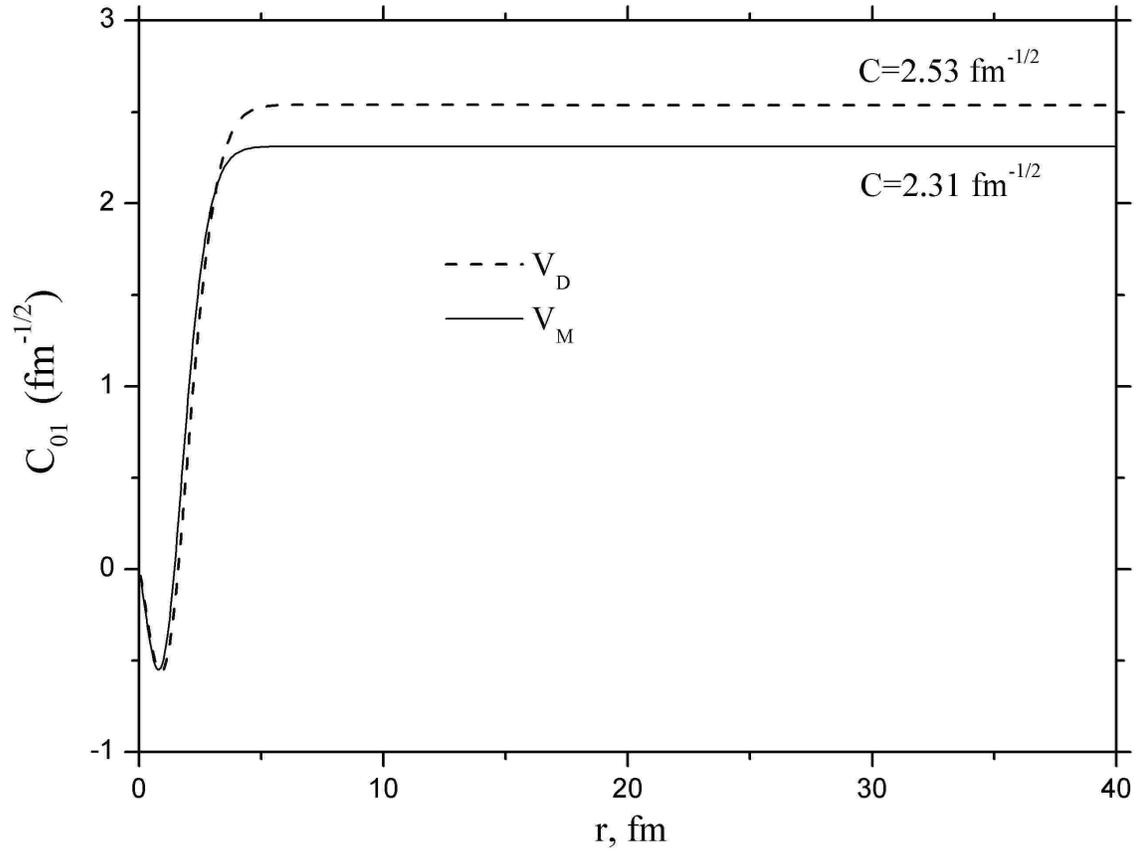}
\caption{Asymptotics of the wave function of the $\alpha+d$ bound
state in the S-wave, calculated with the potentials $V_D$ and
$V_M$.}
\end{figure}

\newpage
%
\begin{figure}
\includegraphics[width=18cm]{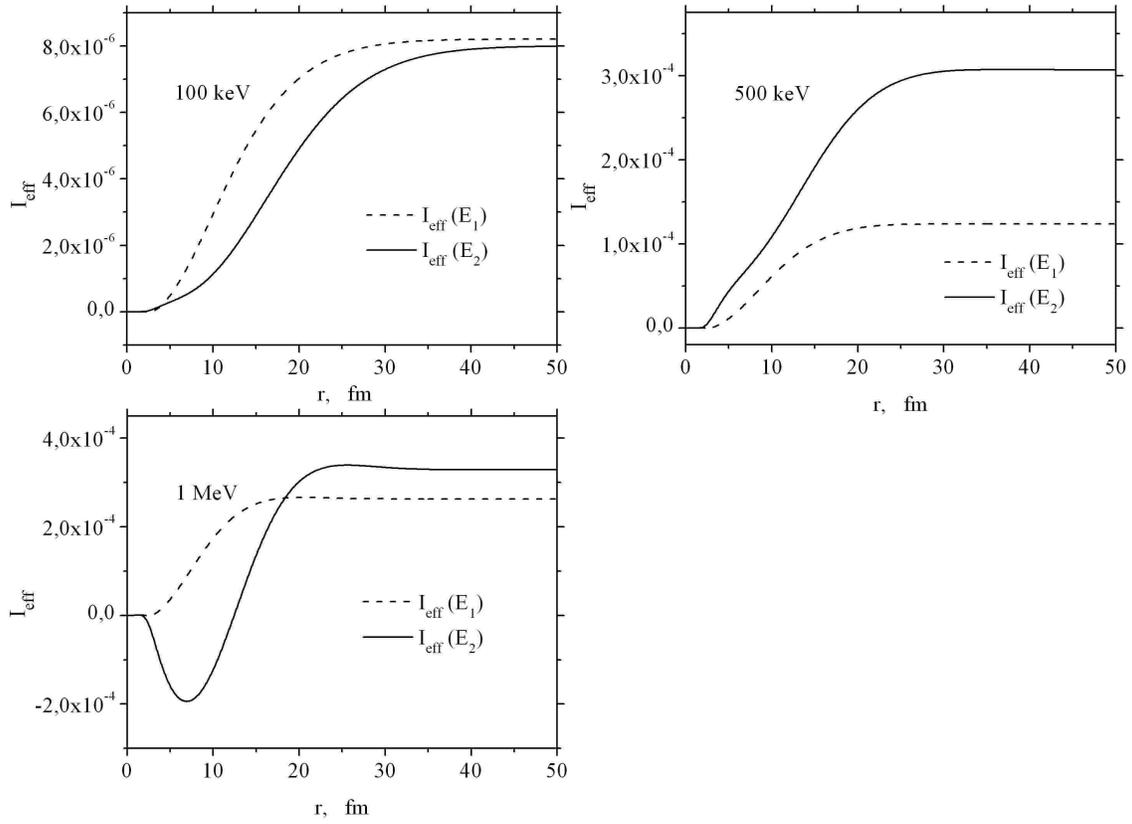}
\caption{Convergence of the effective integrals $I_{eff} $ for
E1($^3P_2->S $)- and  E2($^3D_3->S $)- transitions at energies E=100
keV, E=500 keV, and E=1 MeV.}
\end{figure}
\newpage
%
\begin{figure}
\includegraphics[width=20cm]{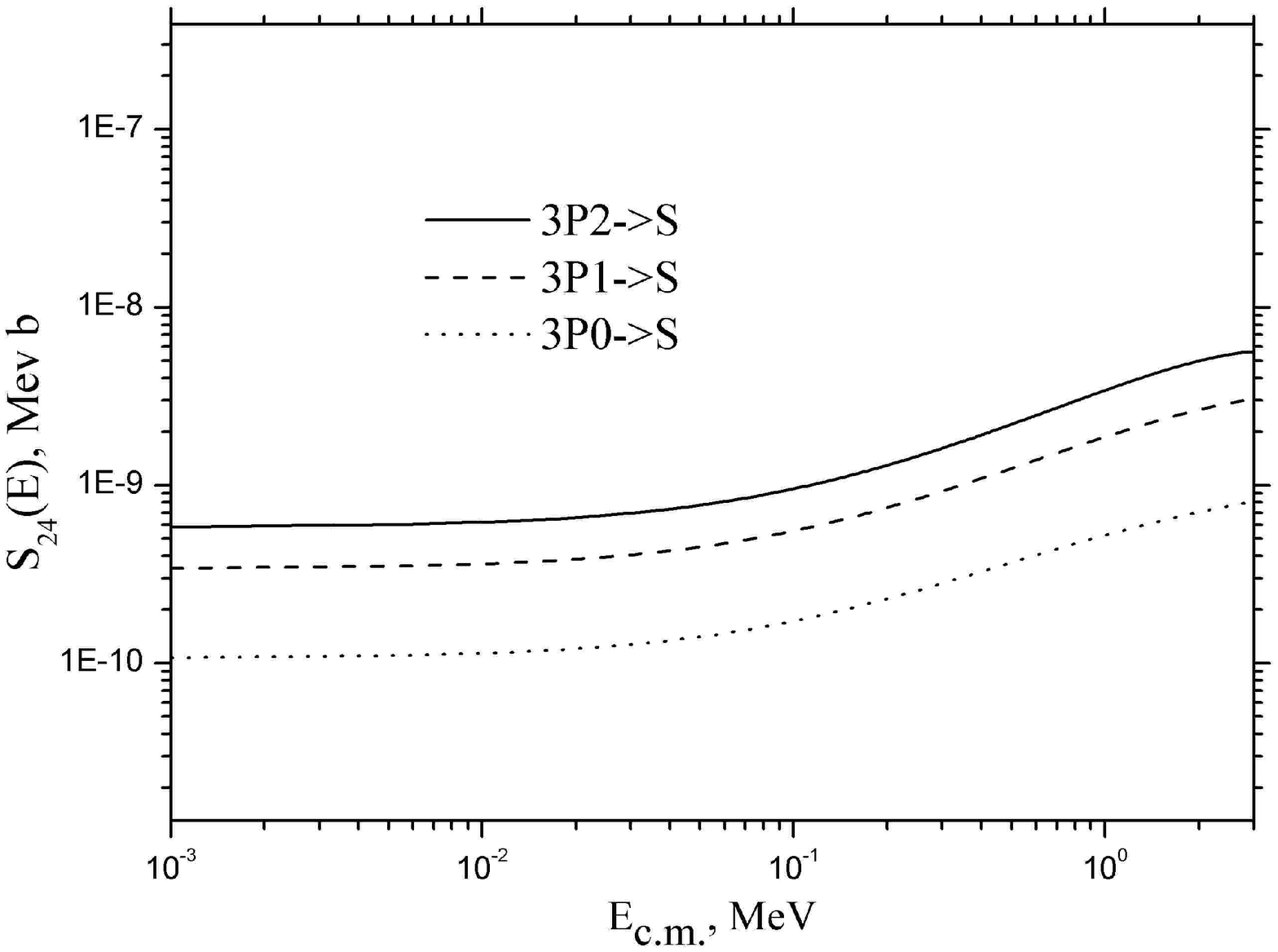}
\caption{Contributions of the E1- components to the astrophysical
S-factor for the synthesis reaction
$\alpha+d\rightarrow^{6}Li+\gamma $ calculated with the potential
 $V_M$.}
\end{figure}

\newpage
%
\begin{figure}
\includegraphics[width=20cm]{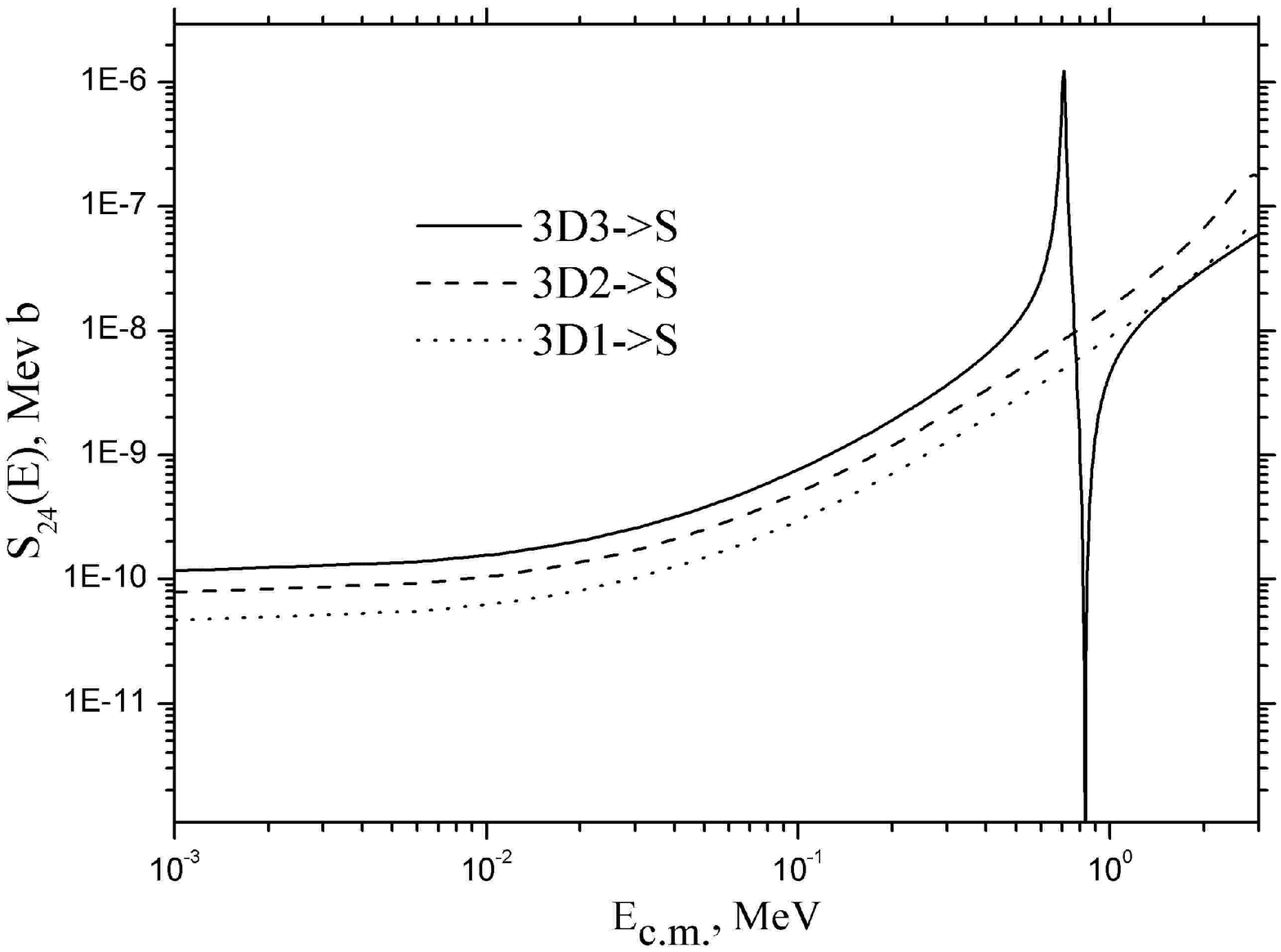}
\caption{Contributions of the E2- components to the astrophysical
S-factor for the synthesis reaction
$\alpha+d\rightarrow^{6}Li+\gamma $ calculated with the potential
$V_M$.}
\end{figure}
\newpage
%
\begin{figure}
\includegraphics[width=20cm]{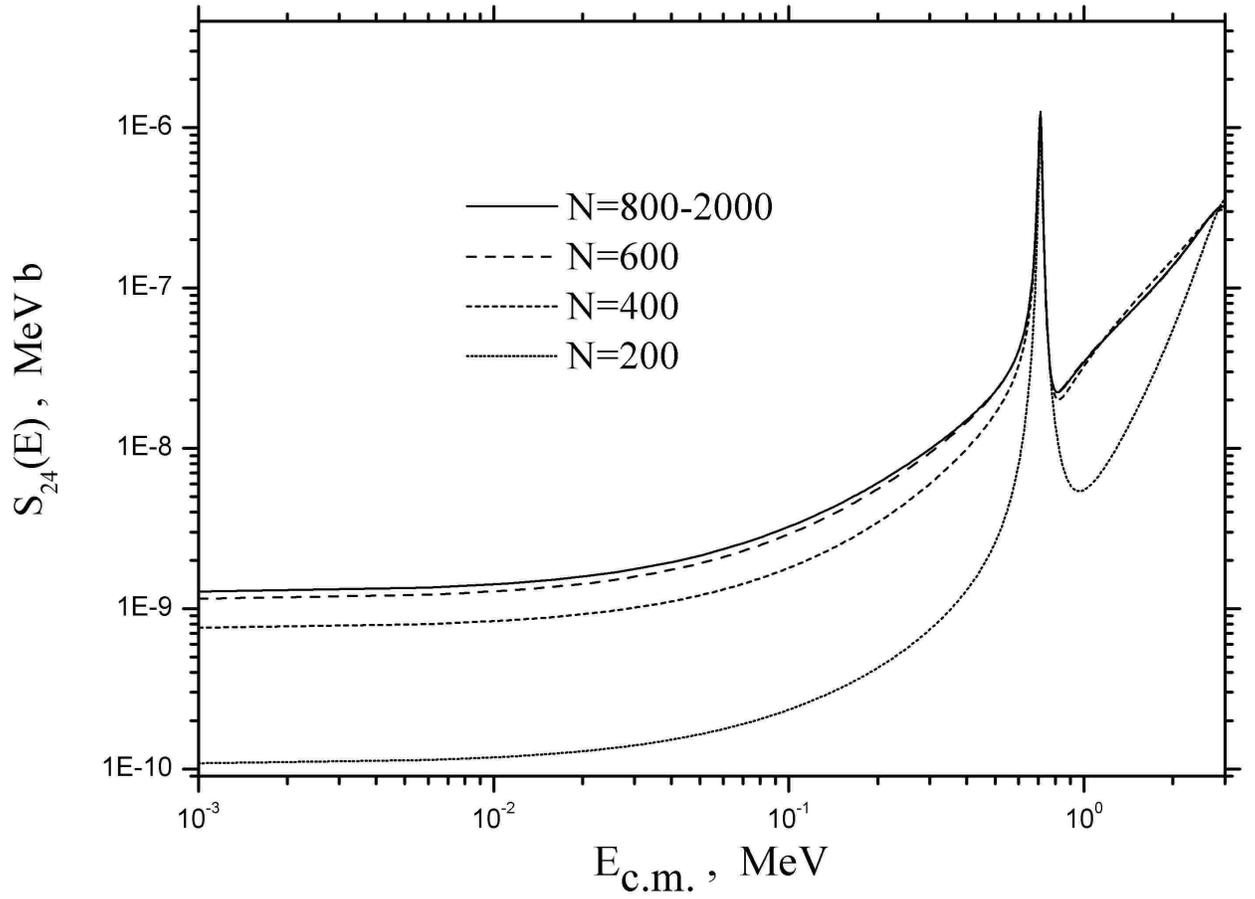}
\caption{Contributions of the E1+E2 transitions to the astrophysical
S-factor for the synthesis reaction
$\alpha+d\rightarrow^{6}Li+\gamma $ calculated with the potential
$V_M$ for different values of the mesh number N (convergence).}
\end{figure}
\newpage
%
\begin{figure}
\includegraphics[width=20cm]{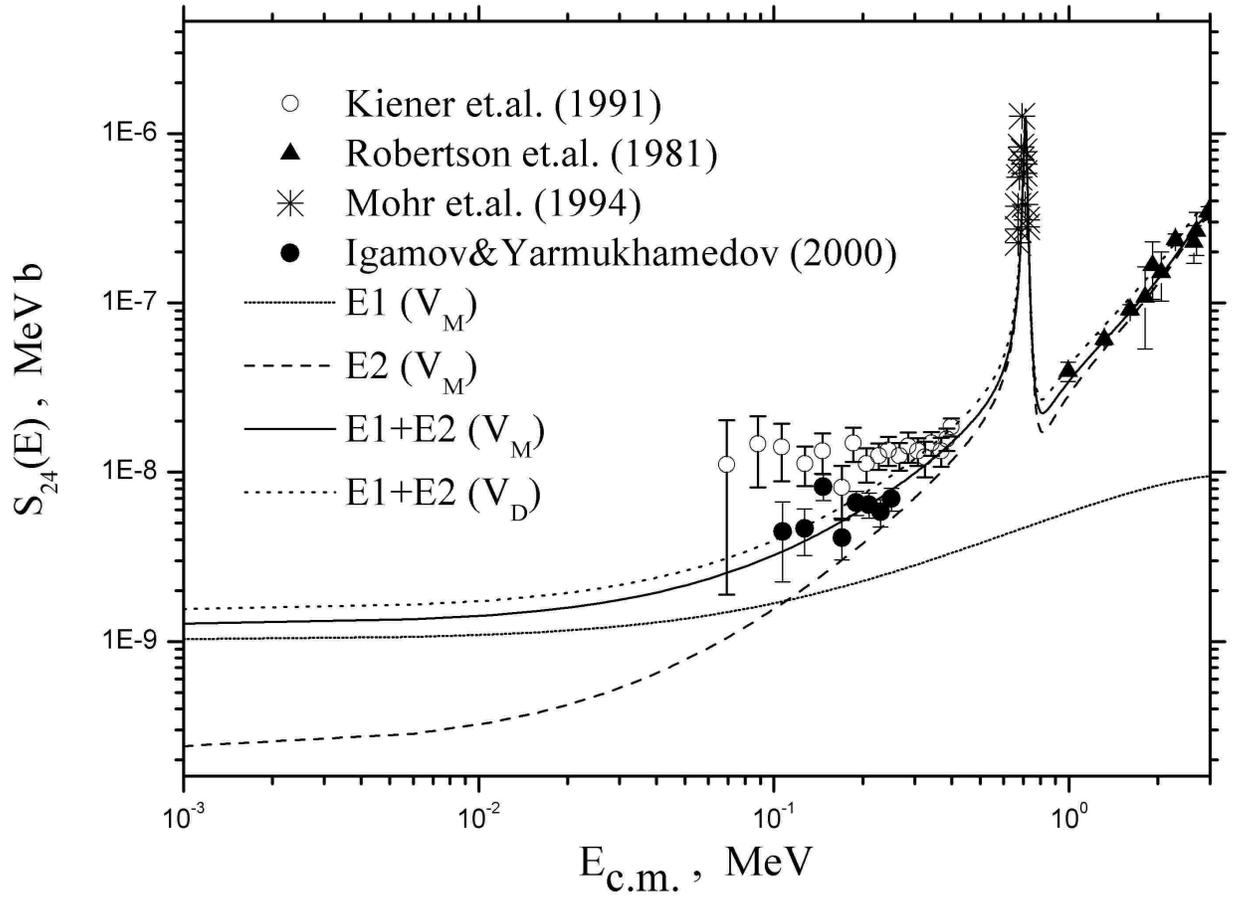}
\caption{Contributions of the E1, E2, E1+E2 transitions to the
astrophysical S-factor for the synthesis reaction
$\alpha+d\rightarrow^{6}Li+\gamma $ calculated with the potentials
$V_M$ and $V_D$ in comparison with the experimental data from
Refs.~\cite{robe81,kien91,mohr94,iga00}.}
\end{figure}

\newpage
%
\begin{table}
\caption {Theoretical estimations for the reaction rates of the
process $d(\alpha,\gamma)^{6}Li$ in the temperature interval $10^{6}
K \leq T \leq 10^{10} K $ ($ 0.001\leq T_{9} \leq 10 $) in
comparison with the results of Refs.~\cite{hamm10,mukh11}.}
\label{tab01}

{\tiny
\begin{tabular}{|c|c|c|c|c|c|}
\hline
$T_9$ &$E_0$  &$N_{a}(\sigma v)$\cite{mukh11} & $N_{a}(\sigma v)$\cite{hamm10} & $N_{a}(\sigma v)$ ($V_D$) & $N_{a}(\sigma v)$ ($V_M$)\\
 & (MeV)    & $( \textrm{cm}^{3} $\rm{mol}$^{-1}$ \rm{c}$^{-1}$ ) & ( \textrm{cm}$^{3}$ \rm{mol}$^{-1} $ \rm{c}$^{-1} )$ &
 $( \textrm{cm}^{3}$ \rm{mol}$^{-1}$ \rm{c}$^{-1}$ ) &
 $( \textrm{cm}^{3}$ \rm{mol}$^{-1}$ \rm{c}$^{-1} )$ \\
  &    & $C_{\alpha d}$=2.30 fm $^{-1/2}$ & $C_{\alpha d}$=2.70 fm $^{-1/2}$ & $C_{\alpha d}$=2.53 fm $^{-1/2}$ & $C_{\alpha d}$=2.31 fm $^{-1/2}$ \\
\hline
0.001 & 0.002 & $6.467\times10^{-30}$ & $9.153\times10^{-30}$ & $6.730\times10^{-30}$& $5.488\times10^{-30}$ \\
0.002 & 0.003 & $1.857\times10^{-23}$ & $2.610\times10^{-23}$ & $2.012\times10^{-23}$& $1.641\times10^{-23}$ \\
0.003 & 0.004 & $2.470\times10^{-20}$ & $3.458\times10^{-20}$ & $2.729\times10^{-20}$& $2.225\times10^{-20}$\\
0.004 & 0.005 & $2.286\times10^{-18}$ & $3.190\times10^{-18}$ & $2.557\times10^{-18}$& $2.085\times10^{-18}$ \\
0.005 & 0.006 & $5.693\times10^{-17}$ & $7.929\times10^{-17}$ & $6.426\times10^{-17}$& $5.241\times10^{-17}$ \\
0.006 & 0.007 & $6.592\times10^{-16}$ & $9.163\times10^{-16}$ & $7.492\times10^{-16}$& $6.110\times10^{-16}$ \\
0.007 & 0.008 & $4.651\times10^{-15}$ & $7.672\times10^{-15}$ & $5.315\times10^{-15}$& $4.334\times10^{-15}$ \\
0.008 & 0.009 & $2.327\times10^{-14}$ & $4.990\times10^{-14}$ & $2.671\times10^{-14}$& $2.179\times10^{-14}$ \\
0.009 & 0.009 & $9.067\times10^{-14}$ & $2.100\times10^{-13}$ & $1.045\times10^{-13}$& $8.520\times10^{-14}$ \\
0.010 & 0.010 & $2.923\times10^{-13}$ & $6.547\times10^{-13}$ & $3.379\times10^{-13}$& $2.755\times10^{-13}$ \\
0.011 & 0.011 & $8.127\times10^{-13}$ & $1.655\times10^{-12}$ & $9.422\times10^{-13}$& $7.684\times10^{-13}$ \\
0.012 & 0.011 & $2.008\times10^{-12}$ & $3.612\times10^{-12}$ & $2.334\times10^{-12}$& $1.904\times10^{-12}$ \\
0.013 & 0.012 & $4.508\times10^{-12}$ & $7.142\times10^{-12}$ & $5.251\times10^{-12}$& $4.282\times10^{-12}$ \\
0.014 & 0.012 & $9.343\times10^{-12}$ & $1.325\times10^{-11}$ & $1.091\times10^{-11}$& $8.895\times10^{-12}$ \\
0.015 & 0.013 & $1.811\times10^{-11}$ & $2.363\times10^{-11}$ & $2.119\times10^{-11}$& $1.728\times10^{-11}$ \\
0.016 & 0.014 & $3.318\times10^{-11}$ & $4.103\times10^{-11}$ & $3.887\times10^{-11}$& $3.170\times10^{-11}$ \\
0.018 & 0.015 & $9.676\times10^{-11}$ & $1.157\times10^{-10}$ & $1.137\times10^{-10}$& $9.273\times10^{-11}$ \\
0.020 & 0.016 & $2.432\times10^{-10}$ & $2.965\times10^{-10}$ & $2.865\times10^{-10}$& $2.336\times10^{-10}$ \\
0.025 & 0.018 & $1.538\times10^{-09}$ & $2.014\times10^{-09}$ & $1.822\times10^{-09}$& $1.486\times10^{-09}$ \\
0.030 & 0.021 & $6.277\times10^{-09}$ & $8.452\times10^{-09}$ & $7.462\times10^{-09}$& $6.085\times10^{-09}$ \\
0.040 & 0.025 & $4.870\times10^{-08}$ & $6.594\times10^{-08}$ & $5.823\times10^{-08}$& $4.749\times10^{-08}$ \\
0.050 & 0.029 & $2.093\times10^{-07}$ & $2.827\times10^{-07}$ & $2.512\times10^{-07}$& $2.049\times10^{-07}$ \\
0.060 & 0.033 & $6.375\times10^{-07}$ & $8.598\times10^{-07}$ & $7.672\times10^{-07}$& $6.257\times10^{-07}$ \\
0.070 & 0.036 & $1.554\times10^{-06}$ & $2.094\times10^{-06}$ & $1.874\times10^{-06}$& $1.528\times10^{-06}$ \\
0.080 & 0.040 & $3.245\times10^{-06}$ & $4.372\times10^{-06}$ & $3.921\times10^{-06}$& $3.198\times10^{-06}$ \\
0.090 & 0.043 & $6.057\times10^{-06}$ & $8.156\times10^{-06}$ & $7.327\times10^{-06}$& $5.977\times10^{-06}$ \\
0.100 & 0.046 & $1.038\times10^{-05}$ & $1.397\times10^{-05}$ & $1.257\times10^{-05}$& $1.026\times10^{-05}$ \\
0.110 & 0.049 & $1.665\times10^{-05}$ & $2.240\times10^{-05}$ & $2.018\times10^{-05}$& $1.646\times10^{-05}$ \\
0.120 & 0.052 & $2.533\times10^{-05}$ & $3.406\times10^{-05}$ & $3.072\times10^{-05}$& $2.506\times10^{-05}$ \\
0.130 & 0.055 & $3.690\times10^{-05}$ & $4.959\times10^{-05}$ & $4.476\times10^{-05}$& $3.651\times10^{-05}$ \\
0.140 & 0.057 & $5.185\times10^{-05}$ & $6.967\times10^{-05}$ & $6.292\times10^{-05}$& $5.133\times10^{-05}$ \\
0.150 & 0.060 & $7.071\times10^{-05}$ & $9.495\times10^{-05}$ & $8.582\times10^{-05}$& $7.001\times10^{-05}$ \\
0.160 & 0.063 & $9.398\times10^{-04}$ & $1.261\times10^{-04}$ & $1.141\times10^{-04}$& $9.307\times10^{-05}$ \\
0.180 & 0.068 & $1.559\times10^{-04}$ & $2.090\times10^{-04}$ & $1.892\times10^{-04}$& $1.543\times10^{-04}$ \\
0.200 & 0.073 & $2.416\times10^{-04}$ & $3.237\times10^{-04}$ & $2.932\times10^{-04}$& $2.392\times10^{-04}$ \\
0.250 & 0.084 & $5.868\times10^{-04}$ & $7.846\times10^{-04}$ & $7.112\times10^{-04}$& $5.805\times10^{-04}$ \\
0.300 & 0.096 & $1.167\times10^{-03}$ & $1.557\times10^{-03}$ & $1.412\times10^{-03}$& $1.153\times10^{-03}$ \\
0.350 & 0.106 & $2.040\times10^{-03}$ & $2.715\times10^{-03}$ & $2.461\times10^{-03}$& $2.010\times10^{-03}$ \\
0.400 & 0.116 & $3.256\times10^{-03}$ & $4.325\times10^{-03}$ & $3.916\times10^{-03}$& $3.199\times10^{-03}$ \\
0.500 & 0.134 & $6.930\times10^{-03}$ & $9.169\times10^{-03}$ & $8.258\times10^{-03}$& $6.752\times10^{-03}$ \\
0.600 & 0.152 & $1.271\times10^{-02}$ & $1.674\times10^{-02}$ & $1.484\times10^{-02}$& $1.215\times10^{-02}$ \\
0.700 & 0.168 & $2.148\times10^{-02}$ & $2.813\times10^{-02}$ & $2.414\times10^{-02}$& $1.979\times10^{-02}$ \\
0.800 & 0.184 & $3.462\times10^{-02}$ & $4.502\times10^{-02}$ & $3.816\times10^{-02}$& $3.144\times10^{-02}$ \\
0.900 & 0.199 & $5.385\times10^{-02}$ & $6.944\times10^{-02}$ & $6.213\times10^{-02}$& $5.191\times10^{-02}$ \\
1.000 & 0.213 & $8.079\times10^{-02}$ & $1.033\times10^{-01}$ & $9.209\times10^{-02}$& $7.755\times10^{-02}$ \\
1.500 & 0.279 & $3.508\times10^{-01}$ & $4.350\times10^{-01}$ & $3.840\times10^{-01}$& $3.312\times10^{-01}$ \\
2.000 & 0.338 & $7.854\times10^{-01}$ & $9.623\times10^{-01}$ & $8.456\times10^{-01}$& $7.342\times10^{-01}$ \\
2.500 & 0.393 & $1.268\times10^{+00}$ & $1.549\times10^{+00}$ & $1.356\times10^{+00}$& $1.177\times10^{+00}$ \\
3.000 & 0.443 & $1.745\times10^{+00}$ & $2.132\times10^{+00}$ & $1.858\times10^{+00}$& $1.609\times10^{+00}$ \\
4.000 & 0.537 & $2.673\times10^{+00}$ & $3.280\times10^{+00}$ & $2.839\times10^{+00}$& $2.438\times10^{+00}$ \\
5.000 & 0.623 & $3.631\times10^{+00}$ & $4.476\times10^{+00}$ & $3.895\times10^{+00}$& $3.321\times10^{+00}$ \\
6.000 & 0.704 & $4.645\times10^{+00}$ & $5.754\times10^{+00}$ & $5.056\times10^{+00}$& $4.291\times10^{+00}$ \\
7.000 & 0.780 & $5.689\times10^{+00}$ & $7.088\times10^{+00}$ & $6.271\times10^{+00}$& $5.309\times10^{+00}$ \\
8.000 & 0.853 & $6.725\times10^{+00}$ & $8.438\times10^{+00}$ & $7.501\times10^{+00}$& $6.342\times10^{+00}$ \\
9.000 & 0.922 & $7.723\times10^{+00}$ & $9.773\times10^{+00}$ & $8.707\times10^{+00}$& $7.354\times10^{+00}$ \\
10.00 & 0.989 & $8.664\times10^{+00}$ & $1.107\times10^{+01}$ & $9.864\times10^{+00}$& $8.325\times10^{+00}$ \\
\hline
\end{tabular}
}
\end{table}

\end{document}